\documentclass[10pt,onecolumn]{IEEEtran} 

\ifCLASSINFOpdf
 \usepackage[pdftex]{graphicx}
\else
  \usepackage[dvips]{graphicx}
\fi

\usepackage[cmex10]{amsmath}
\usepackage{bbm}
\usepackage{amssymb}
\usepackage{dsfont}

\usepackage[font={small}]{caption}
\usepackage{subcaption}

\usepackage{graphicx}
\usepackage{color}
\usepackage{transparent}

\usepackage{tikz}
\usetikzlibrary{patterns}
\usepackage{pgf}
\usepackage{pgfplots}
\usepackage{pgfplotstable}
\pgfplotsset{compat=newest}
\usepgfplotslibrary{patchplots}

\usepackage{url}

\newtheorem{defi}{Definition}
\begin{document}

\title{Managing Flexibility in Multi-Area Power Systems}

\author{\IEEEauthorblockN{Matthias A. Bucher, Spyros Chatzivasileiadis and G\"{o}ran Andersson\\}
\IEEEauthorblockA{Power Systems Laboratory, ETH Zurich, Switzerland, \{mb, spyros, andersson\}@eeh.ee.ethz.ch\\}}

\maketitle
\begin{abstract}
In this paper we present a framework to efficiently characterize the available operational flexibility in a multi-area power system. We focus on the available reserves and the tie-line flows. The proposed approach is an alternative to the current calculation of the Available Transfer Capacity (ATC), as it considers location and availability of reserves, transmission constraints, and interdependencies of tie-line flows between different areas, while it takes into account the N-1 security criterion.
The method is based on computational geometry using polytopic projections. It requires only a limited amount of information exchange and does not need central coordination. The method has two versions: a passive and an active approach, where neighboring areas can share reserves. In that respect we also introduce the term ``exported flexibility'', which could form the basis for a new trading product in electricity markets. Case studies demonstrate the improved tie-line utilization, especially if reserves are shared, and the visualization benefits.
\end{abstract}

\IEEEpeerreviewmaketitle

\section{Introduction}

Power systems in Europe, as well as in other parts of the world, are undergoing fundamental changes. First, increasing shares of fluctuating renewable energy sources are connected to the network. Second, electricity markets are becoming larger covering wider areas. Since in Europe regional markets are merged towards a common European energy exchange\cite{IEMDirective}.
These developments result in more frequent power flow changes and higher power transfers which stretch over longer distances. Substantial network reinforcements are often essential to accommodate such flows and ensure power system security \cite{ENTSOETEN}. However, long licensing procedures for building new lines, increased public opposition and high investment costs call for additional measures to tackle the emerging problems. Besides building additional lines, flexibility in power system can be achieved by fast-reacting power sources, flexible loads, and a better utilization of the transmission line infrastructure. This paper will focus on the computation of the cross-border transmission line limits, to increase power system flexibility and account for the contribution of flexible resources in neighboring areas to deal with contingencies occuring in the area in question. In order to properly account for the interdependencies between cross-border available transmission capacities and the flexible resources in each area, a centralized computation seems as the most straightforward approach. Due to the numerous Transmission System Operators (TSOs), as for example in Europe, the control reserves in different control areas are currently operated mainly independently and without coordination. It should be noted, nevertheless, that first steps towards an improved coordination have been taken in countries such as Germany where the procurement and operation of reserves of the four TSOs has been merged. Furthermore, the establishment of bodies such as TSC, a cooperation initiative between thirteen European TSOs, and CORESO, which acts as a coordination service center with the objective to enhance the level of security of supply in Europe, are further initiatives addressing the aforementioned challenges in the European context.

Numerous definitions of flexibility exist \cite{FlexTaxonomy}. For the scope of this paper we define {\it operational flexibility} as follows:\\

\begin{defi}
{\it The operational flexibility of a system is the ability of the system to react to a disturbance sufficiently fast in order to keep the system secure. A disturbance can either be a component outage, e.g. a line, generator, or a deviation of power injection, e.g. due to forecast errors.\\}
\end{defi}

Main sources of flexibility are generation sources, which are usually contracted to provide regulating energy  (spinning reserves) or are able to be redispatched sufficiently fast (manual reserves). Besides generation however, switching operations, demand response, electricity storage, and the power flow controllability of new components such as HVDC lines can also provide a significant source of flexibility (for HVDC see e.g.~\cite{SCH_Powertech_2013_1}). A metric to quantify available operational flexibility in terms of available energy, power capacity and up- and down ramping is for example introduced by the authors in \cite{Makarov2009}.

Flexibility in a power system's context has been discussed in various publications. For example, Ref.~\cite{PolytopeAAU} presents a polytope-based method in order to represent the operational limits over time. The aggregation of multiple units can be found as a Minkowski summation as long as no transmission constraints are imposed. In \cite{Lannoye2012} a method to estimate the probability of insufficient ramping capabilities is presented and \cite{Kirschen2011} tries to optimize the flexibility of a generation mix.

In this paper we focus on the coordination of flexibility {\it between} TSOs and on inter-TSO flows. 
Envisioned is an operation paradigm, where TSO A communicates the boundaries of allowed power flow deviations on the tie-lines to TSO B and vice-versa, taking into account the available reserves in both areas. TSO B can use this information when it needs additional flexibility: for example, it can deviate from the agreed tie-line flows, knowing that TSO A will be able to act correctively to handle any contingencies occuring in area A from the tie-line flow change. Or, inversely, it can guarantee that it can take the necessary control actions which will ensure a secure operation in TSO A's control area. Both such actions deal with flexibility offered to the neighboring area in order to ensure a secure power system operation. Therefore we introduce and define the term {\it exported flexibility}:\\
\begin{defi}
{\it The exported flexibility, is the operational flexibility originating in a control area that can be used by neighboring control areas.\\}
\end{defi}

In section VI, the exported flexibility will be quantified and compared in different cases.\\


The main contributions of this paper are the following.
First, flexibility metrics in literature are mostly determined for a single control area, focusing mainly on generation, while no transmission constraints are assumed, i.e.``copperplate approach''. We present a method for multiple areas taking into account the inter- and intra-area transmission constraints and their interdependencies.
Second, the approach is decentralized, not requiring any central coordination, i.e. the TSOs exchange the relevant information bilaterally. Therefore, it can be applied to the current operation paradigm. Third, exploiting spatial information about the contracted reserves, i.e. assuming that the TSO knows the location of the contracted reserves in its area, we show that the tie-line utilization can be improved compared with the current usual calculation of the Available Transmission Capacity (ATC). The ATC currently reflects the maximum tie-line flow which can be exchanged between two areas, without leading to any contingencies to any of the two areas. The ATC calculation is conservative, as it has no spatial information about the location of the reserves and always assumes the worst case.\\

The method is based on computational geometry and has several advantages and features. First, through the use of polytopic projections, only the limits on the interconnecting power flows are communicated between the TSOs. No explicit information pertaining to potentially confidential data, such as the grid topology, committed generators or reserve availability need to be exchanged.
Taking further advantage of computational geometry techniques, the approach allows a straightforward visualization of the resulting flexibility metric. 

In the method, we distinguish between two alternatives: an {\it active approach}, where corrective control measures are allowed, and a {\it passive approach}, where no corrective control measures are used.
Both approaches consider spatial information concerning which nodes power will be injected to and the corresponding anticipated power flow changes.
The {\it passive approach}  leads to a set of possible combinations of power flow deviations on the tie-lines without causing congestions.
The {\it active approach} enables reserves sharing between the control areas. Therefore the range of possible tie-line flows is larger, as we assume that if TSO B incurs changes to the tie-line flows between areas A and B, TSO A is able to carry out the necessary corrective control actions in area A to avoid any contingencies. The tie-line utilization for reserve operation compared to current ATC calculation is therefore improved.

The remainder of this paper is structured as follows: In Section II we define the problem. In Section III we formulate the equations describing the available flexibility and in Section IV we build the flexibility sets for the active and passive approach. Section V presents the proposed method for computing the tie-line limits. The performance of this method is investigated in a case study in Section VI. Section VII concludes the paper and gives an outlook of future research.

\newpage
\section{Definition of the Problem and Approach}
\label{sec:problem_def}

\begin{figure}[ht]
\center
  \begin{minipage}[b]{0.25\linewidth}
    \centering
    \includegraphics[width=\linewidth]{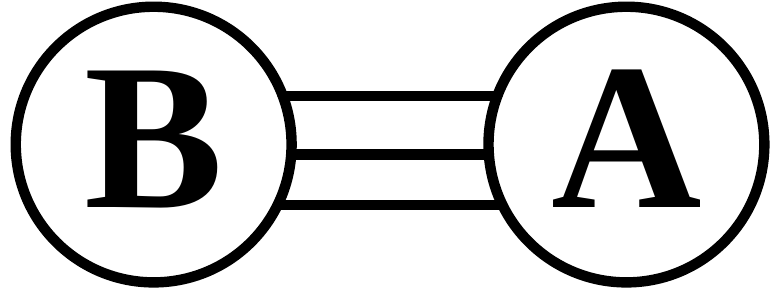}
    \caption{Two area system.}
    \label{fig:multiareaabstract}
  \end{minipage}
  \hspace{0.2cm}
  \begin{minipage}[b]{0.25\linewidth}
    \centering
    \includegraphics[width=\linewidth]{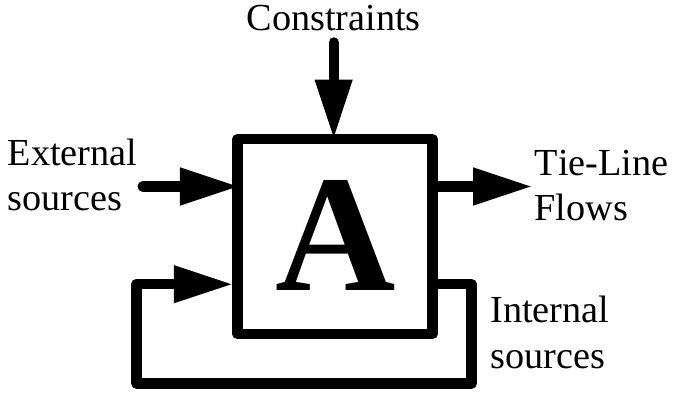}
    \caption{MIMO System.}
    \label{fig:mimo}
  \end{minipage}
\end{figure}

In this paper, we consider a power system, divided in two control areas where tie-lines (AC and DC) form the interconnection between the two neighboring areas. For brevity and clarity we focus on two control areas. The generalization to more than two areas is subject of future work.
Fig. \ref{fig:multiareaabstract} shows the two control areas $A$ and $B$. Each area is assumed to be controlled by an individual TSO. The power flows on the tie-lines are influenced by the unit commitment and dispatch in both regions.

The proposed method should provide the limits of flexible resources in area A which could assist area B in case of a disturbance, and vice-versa.

As shown in Fig. \ref{fig:mimo} control area A can be further abstracted as a system with multiple inputs and multiple outputs (MIMO system). We distinguish between {\it external sources} and {\it internal sources}. External sources are inputs that can directly be influenced by neighboring areas such as flows on tie-line or HVDC interconnections. Internal sources are controlled locally, e.g. generation units within area A. This MIMO system is subject to constraints, such as intra-area transmission capacity limits and generation capacities.\\

Ultimately, the goal of this work is to find the set of allowed setpoint combinations of external sources such that all the operational constraints are satisfied.

\section{Quantification of Flexibility}

The available flexibility depends on the current system state, i.e. the dispatch. We assume that the dispatch is given, i.e. the generation units are injecting active power at setpoints given by the vector $P_{gen}$ such that the demand $P_{load}$ is covered and operational constraints are respected.
The net injections per bus $P_{Bus} = P_{gen} - P_{load}$ lead to power flows $P_L$ that have to remain within transmission limits $\left[\underline{P}_L, \overline{P}_L \right]$.

This could be considered as the outcome of a market operation. In order to guarantee a secure operation, the N-1 security criterion is commonly used to make sure that the outage of a single element does not lead to a blackout. The given dispatch is required to fulfill the N-1 criterion.

\subsection{Power Flows per Area and $\Delta$-Notation}

We split the system into two control areas as in Section \ref{sec:problem_def} and distinguish for every control area the contribution to the power flows caused by internal sources and external sources (Eq. \eqref{eq:powerflow}). We use the PTDF (Power Transfer Distribution Factor) matrix in order to determine the influence of changes in the bus injections from internal and external sources on the line power flows.

The PTDF matrix describes the sensitivity $H_{jb}$ of a specific power flow $P_{L,j}$ on line $j$ with respect to a change of the net power injection $P_{Bus,b}$, i.e. the difference between total generation and the total load at a bus $b$. In order to maintain the active power balance, the corresponding power is extracted at the reference bus \cite{Wollenberg2000}.
The PTDF matrix $H$ for the considered area can be split into $\left[ H_i \ H_e \right]$ where the indices $i$ denote the influences of $n_i$ internal sources and $e$ for $n_e$ external sources on the power flows of the considered area.
The net bus injections $P_{Bus,i}$ can be split into injections from internal and injections from external sources: $ P_{Bus} = \left[ P_{Bus,i}\ P_{Bus,e} \right]^T$ and further, they can be split into the scheduled power injections $P_{Bus,sched}$ and deviations $p$, i.e. $P_{Bus} = P_{Bus}^{sched} + p$.

The power flows can be split into scheduled flows $P_{L}^{sched}$ and unscheduled flows $P_{L}^{dev}$ and can be assigned to be caused by an {\it internal} or {\it external} source:

\begin{equation}
\begin{aligned}
P_{L} = \underbrace{H_i P_{Bus,i}^{sched} + H_e P_{Bus,e}^{sched}}_{= P_{L}^{sched}} + \underbrace{H_e p_e + H_i p_i}_{= P_{L}^{dev}} \\
\end{aligned}
\label{eq:powerflow}
\end{equation}

For a given generation dispatch, we can quantify the remaining transmission capacity for each line until a constraint becomes binding. The remaining transmission capacity is calculated as the difference to the capacity, i.e. $\Delta\underline{P}_{L} = \underline{P}_{L} - P_{L}^{sched}$ and $\Delta\overline{P}_{L} = \overline{P}_{L} - P_{L}^{sched}$ respectively.

The allowed changes in the power output at every bus are given by the allowed changes to the schedule, i.e. $\Delta\underline{P}_{Bus} = P_{res}^{dn}$ and $\Delta\overline{P}_{Bus} = P_{res}^{up}$ respectively. $P_{res}^{dn},P_{res}^{up}$ are vectors with the maximum changes to the operation setpoint of the generators that the TSO is allowed to make at every bus. These changes could be available manual reserves or redispatch measures.
The remaining up- or down ramping capabilities are subject to ramping rates and start-up times constraints and thus the amount of power that can be ramped up or down is additionally limited by the given time to ramp. We do not consider ramp rates in this work. The incorporation will be the topic of future work. \\

For the remainder of this paper, we only consider the deviations of the setpoints in the vectors $p_i,p_e$ and the corresponding limits written in the $\Delta$-notation, i.e. $\Delta\underline{P}_{L}$, $\Delta\overline{P}_{L}$, $\Delta\underline{P}_{Bus}$ and $\Delta\overline{P}_{Bus}$.

In the next step we quantify the flexibility by formulating the limits on possible combinations of deviations of internal and external sources. The resulting inequality constraints will be combined in a set that we refer to as {\it flexibility set}.

\subsection{Constraints for nominal operation}

We consider first the constraints that have to be satisfied during normal operation, i.e. without any outages. We refer to this as the {\it N-secure} case. The deviations of the net injections have to be within the allowed changes of the power flow on the transmission lines:
\begin{equation}
\begin{aligned}
\Delta \underline{P}_{L} \leq & \begin{bmatrix}
H_i & H_e
\end{bmatrix}
\begin{bmatrix}
p_i \\
p_e
\end{bmatrix}
\leq \Delta \overline{P}_{L} \\
\end{aligned}
\label{eq:constraints1}
\end{equation}

The changes of the bus injections have to meet the generation capacities. $I$ is a unity matrix.
\begin{equation}
\begin{aligned}
\Delta \underline{P}_{bus} \leq & \begin{bmatrix} I_{i} & 0 \\
0 & I_{e}
\end{bmatrix}
\begin{bmatrix} p_i \\ p_e \end{bmatrix} \leq \Delta \overline{P}_{bus} \\
\end{aligned}
\label{eq:constraints2}
\end{equation}

The power balance has to be fulfilled, i.e. the sum of the deviations has to be zero. We can rewrite every equality constraint by two inequality constraints. $\mathbf{1}$ is a row vector of ones.

\begin{equation}
\left[
\mathbf{1}_{1\times n_i} \ \mathbf{1}_{1\times n_e}
\right]
\begin{bmatrix}
p_i \\
p_e
\end{bmatrix}
= 0 \Leftrightarrow
0 \leq \left[ \mathbf{1}_{1\times n_i} \ \mathbf{1}_{1\times n_e} \right]
\begin{bmatrix}
p_i \\
p_e
\end{bmatrix}
\leq 0, \\
\label{eq:constraints3}
\end{equation}


\subsection{N-1 security constraints}
In the N-1 secure case we consider additional constraints for every single line outage and generation unit outage using the GGDF \cite{GGDF} and LODF \cite{Wollenberg2000} matrices. GGDF (Generalized Generation Distribution Factor) gives the changes of the line power flows for an outage of a generation unit $P_{gen,k}$. In this paper, it is assumed that the lack of power is distributed on the remaining generators relative to their total capacity. This could be interpreted as the obligation of primary control reserve provision. $G$ is the GGDF matrix where $G_{jk} P_{gen,k}$ is the change on line $j$ for the outage of generation unit $k$ that produced $P_{gen,k}$ before \cite{GGDF}.
The LODF (Line Outage Distribution Factor) matrix gives the changes of power flows on the lines in the case of a line outage. $L$ is the LODF matrix where $L_{jh} P_{L,h}$ is the change of line flows on line $j$ when line $h$, carrying $P_{L,h}$, trips \cite{Wollenberg2000}.
This adds additional constraints for every outage, but the number of constrained variables is not increased. Usually most of the constraints are not binding and can thus be removed.

We consider every possible outage in the set $\mathcal{G}$ of every single generator $k$.

\begin{equation}
\begin{aligned}
& Q_k
\begin{bmatrix}
p_i \\
p_e
\end{bmatrix}
+ G_k P_{gen,k}
\leq \Delta \overline{P}_{L} \ \ \ \forall k \in \mathcal{G}\\
& Q_k
\begin{bmatrix}
p_i \\
p_e
\end{bmatrix}
- G_k P_{gen,k}
\geq \Delta \underline{P}_{L}  \ \ \ \forall k \in \mathcal{G}\\
\end{aligned}
\label{eq:busout}
\end{equation}

Further, the line outages in the set $\mathcal{L}$ are considered as:

\begin{equation}
\begin{aligned}
& R_j \begin{bmatrix}
p_i \\
p_e
\end{bmatrix}
+ L_j P_{L,j}^{sched}\leq \Delta \overline{P}_{L}\ \ \ \forall j \in \mathcal{L}\\
& R_j \begin{bmatrix}
p_i \\
p_e
\end{bmatrix}
- L_j P_{L,j}^{sched}
\geq \Delta \underline{P}_{L} \ \ \ \forall j \in \mathcal{L}\\
\end{aligned}
\label{eq:lineout}
\end{equation}

The matrices $Q$ and $R$ are defined as:

\begin{equation}
\begin{aligned}
Q_k &= \begin{bmatrix}
Q_{i,k} & Q_{e,k}
\end{bmatrix} \\
&=\begin{bmatrix}
H_{i,\{c <k \}} & H_{i,\{c = k\}}+G_k & H_{i,\{c > k\}} & H_e
\end{bmatrix}\\
R_j &= \begin{bmatrix}
R_{i,j} & R_{e,j}
\end{bmatrix}
=L_{j} \begin{bmatrix}
H_{i,\{r=j\}} & H_{e,\{r=j\}}
\end{bmatrix} \\
\end{aligned}
\end{equation}

Where $c$, $r$ denote the  $c^{\textnormal{th}}$ column and the $r^{\textnormal{th}}$ row  and $G_k$ and $L_j$ are the $k^{\textnormal{th}}$ and $j^{\textnormal{th}}$ column of the corresponding matrix.

\subsection{Representation in Matrix Form}
The constraints Eq. \eqref{eq:constraints1} - \eqref{eq:lineout} can be written as a matrix inequality of the form $C_i p_i + C_e p_e \leq b$. The matrix $C_i$ relates to the internal sources and $C_e$ to external sources. The limits are given by the vector $b$. The constraints in Eq. \eqref{eq:constraints1}-\eqref{eq:constraints3} represent the constraints for nominal operation. The equations can be compiled to
\begin{equation}
\begin{aligned}
C_{i,N} p_i + C_{e,N} p_e \leq b_N.
\end{aligned}
\end{equation}

Analogously, the constraints related to N-1 security (Eq. \eqref{eq:busout},\eqref{eq:lineout}) are stacked to
\begin{equation}
\begin{aligned}
C_{i,N-1} p_i + C_{e,N-1} p_e \leq b_{N-1}.
\end{aligned}
\end{equation}

The matrices $C_{i,N}, C_{e,N}$ and vector $b_N$ are given by:

\begin{equation}
\begin{aligned}
C_{i,N} &= \left[
H_{i}^T ,-H_{i}^T , \left[ I_{i}\ 0 \right]^T , \left[ -I_{i}\  0\right]^T,\mathbf{1}_{n_i \times 1} ,\mathbf{1}_{n_i \times 1}
\right]^T\\
C_{e,N} &= \left[
H_{e}^T ,-H_{e}^T , \left[ 0 \ I_{e} \right]^T, \left[0\ -I_{e} \right]^T ,\mathbf{1}_{n_e \times 1} ,\mathbf{1}_{n_e \times 1}
\right]^T\\
b_{N} &=\left[
\Delta \overline{P}_{L}^T ,-\Delta \underline{P}_{L}^T ,\Delta \overline{P}_{bus}^T ,-\Delta \underline{P}_{bus}^T ,0 ,0
\right]^T\\
\end{aligned}
\end{equation}

The matrices $C_{i,N-1}, C_{e,N-1}$ and vector $b_{N-1}$ are given by:
\begin{equation}
\begin{aligned}
&\forall k \in \mathcal{G}, j \in \mathcal{L}:\\
C_{i/e,N-1} &=\left[
Q_{i/e,k}^T ,-Q_{i/e,k}^T ,R_{i/e,j}^T ,-R_{i/e,j}^T
\right]^T \\
b_{N-1} &=\left[ (\Delta \overline{P}_{L}  - G_k P_{gen,k})^T ,(-\Delta \underline{P}_{L} - G_k P_{gen,k})^T, \right. \\
& \left. \ \ \ \ (\Delta \overline{P}_{L} - L_j P_{L,j}^{sched})^T,(-\Delta \underline{P}_{L} - L_j P_{L,j}^{sched})^T \right]^T \\
\end{aligned}
\end{equation}

\newpage
\section{Flexibility Sets}
The matrix inequalities above defines a set which describes all possible combinations of deviations that are feasible. This set is referred to as the {\it flexibility set}.
We formulate the flexibility set for an {\it active} and a {\it passive} approach as well as for the N and the N-1 secure case, as will be shown below.

For the flexibility set of the {\it active approach}, we allow generation units to adapt their operation point, i.e. the TSO uses the flexibility at his disposal in order to react on tie-line flow changes. For the flexibility set of the {\it passive approach} the TSO does not react on tie-line flow changes.

\subsection{Flexibility Set for Active Approach}

The deviations of external sources together with a corresponding reaction of internal sources that are feasible form the {\it active flexibility set}.

\noindent In the nominal case ({\it Active/N}):

\begin{equation}
\begin{aligned}
F =\{ &\left( p_i, p_e \right) \in \mathbb{R}^{n_i} \times \mathbb{R}^{n_e} \vert
& C_{i,N} p_i + C_{e,N} p_e \leq b_N \} \\
\end{aligned}
\end{equation}

\noindent In the N-1 secure case ({\it Active/N-1}):

\begin{equation}
\begin{aligned}
F = \{ &\left( p_i, p_e \right) \in \mathbb{R}^{n_i} \times \mathbb{R}^{n_e} \vert
\begin{bmatrix}
C_{i,N} \\ C_{i,N-1}
\end{bmatrix} p_i +
\begin{bmatrix}
C_{e,N} \\ C_{e,N-1}
\end{bmatrix} p_e \leq \begin{bmatrix}
b_N \\ b_{N-1}
\end{bmatrix} \} \\
\end{aligned}
\end{equation}

As the N-1 criterion is incorporated via Eqs. \eqref{eq:busout},\eqref{eq:lineout}, the allowed deviations are constrained such that an outage in the considered region can happen without causing N-1 violations.

\subsection{Flexibility Set for Passive Approach}
A special case of the active approach is when $p_i = 0$ and thus the internal sources are not deviating from the setpoints, i.e. the generation units are producing according to their initial dispatch. We call this special case the passive approach, as the resulting feasible deviations of the external sources don't lead to any local congestions.

\noindent By setting $p_i = 0$, the {\it passive flexibility set} in the nominal case ({\it Passive/N}) becomes:
\begin{equation}
\begin{aligned}
& F = \{ p_e  \in \mathbb{R}^{n_e} \vert C_{e,N} p_e \leq b_N \} \\
\end{aligned}
\end{equation}

\noindent In the N-1 secure case ({\it Passive/N-1}):

\begin{equation}
\begin{aligned}
F = \{ &p_e \in \mathbb{R}^{n_e} \vert \begin{bmatrix}
C_{e,N} \\ C_{e,N-1}
\end{bmatrix} p_e \leq \begin{bmatrix}
b_N \\ b_{N-1}
\end{bmatrix} \} \\
\end{aligned}
\end{equation}

\noindent This defines sufficient margins on the transmission lines in the case of an outage.

\newpage
\section{Methodology and Applications}

\subsection{Methodology}
\begin{figure}[h]
\center
\includegraphics[width=0.3\linewidth]{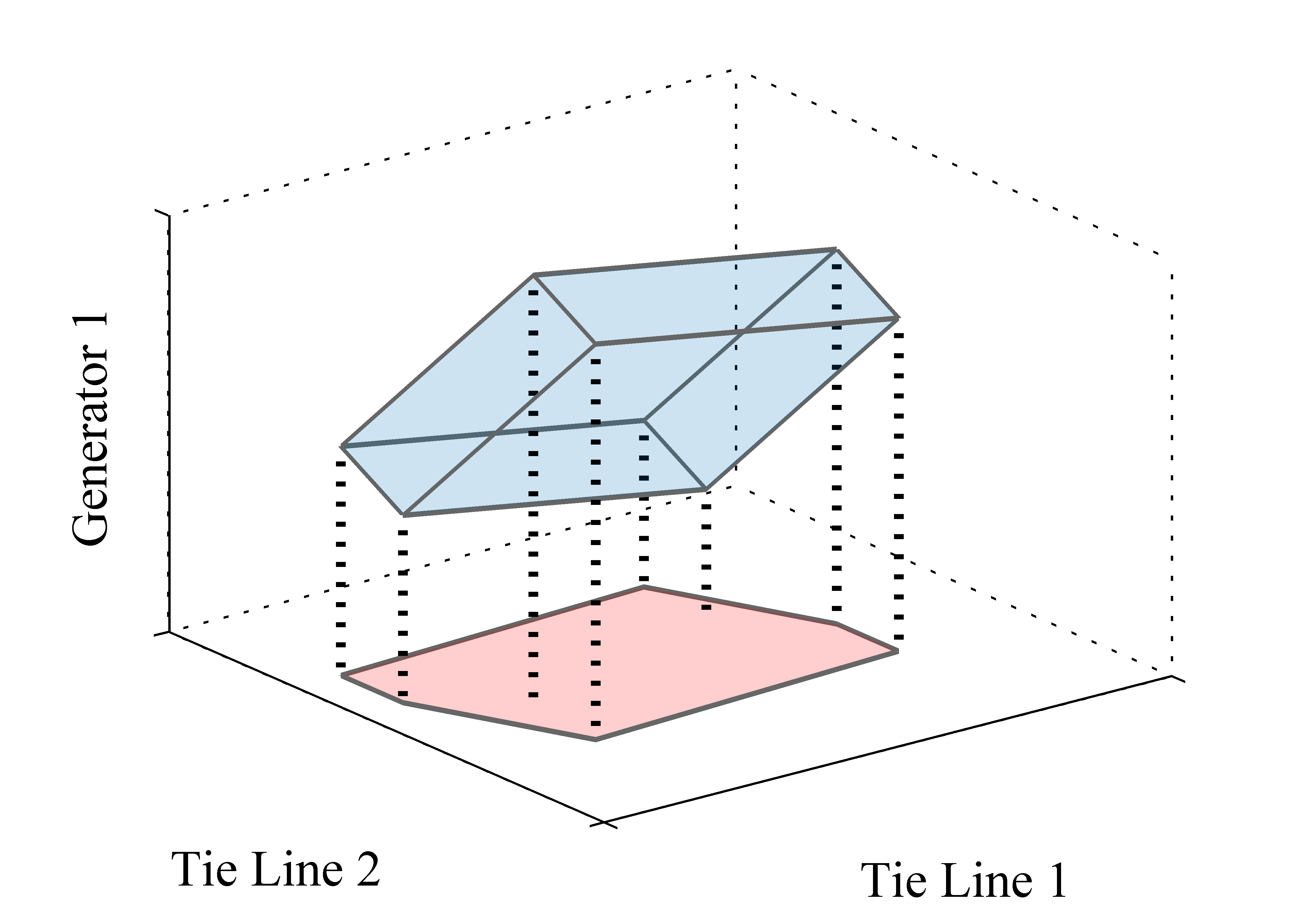}
\caption{The projection represents the feasible set for the external sources.}
\label{fig:example}
\end{figure}

The flexibility set $F$ defines the allowed deviations from the current system state. In order to minimize the information and conceal internal data, the desired outcome is a {\it reduced} set of constraints only depending on the deviations of the tie-line flows, i.e. what combinations of setpoints for the external sources are safe.
Depending on the selected external input, there may be internal adaptation of generation setpoints needed.
In order to illustrate the methodology we consider as an example the flexibility set which is depicted in Fig. \ref{fig:example} as a blue polytope. The system considered consists of two external sources, e.g. two tie-lines, and one internal, e.g. a generator that can be redispatched.

Every feasible setpoint for external sources $F_e$ is then given by the red set, which is the projection of $F$ on the plane spanned by the external sources. Thus, determining $F_{e}$ is equivalent to calculating the projection of $F$ on the dimensions related to the external sources. Or in other words, a setpoint for the two external sources is feasible, if we can adjust the internal sources such that the resulting operating point is within the flexibility set.
In this example we consider only one internal source. In the general case, the number of internal sources $n_i$ corresponds to units that can be redispatched and the number of external sources $n_e$ relates to the number of tie-lines. $F_e$ can be written as:

\begin{equation}
\begin{aligned}
F_{e}   &= \{ p_e \in \mathbb{R}^{n_e} \vert \  \exists \ p_i, \left( p_i,p_e \right) \in F \} = \{ p_e \in \mathbb{R}^{n_e} \vert \ G p_e \leq g \} \\
\end{aligned}
\end{equation}

As long as setpoint deviations are within the set $F_e$ of the passive approach, no congestions occur and thus no redispatch is necessary. For deviations in the set of the active approach, which are not in the passive set, the TSO has to adjust the setpoints of selected generation units with respect to the violated operational constraint.

The outcome of the projection is a linear matrix inequality $G p_e \leq g$ which spans the set of allowed combinations of deviations of the external sources. The number of rows of $G$ is problem dependent and the number of columns is $n_e$. This linear matrix inequality is communicated to the neighboring TSO.
The information gives the neighboring TSO the bounds of possible tie-line flow deviations but does not disclose potentially confidential data to the neighboring TSO, e.g. the generation dispatch of the local control area.
The projection can be performed by known algorithms such as the Fourier-Motzkin-Elimination or the Equality Set Projection \cite{JonesESP}.  
We use the Multi-Parametric Toolbox \cite{MPT3} for the calculation of the projections. The calculations presented in the case study are done within a computing time of around 10 seconds on a standard desktop computer. Future work will investigate the factors influencing the computing time as well possibilities to reduce the problem complexity.

\subsection{Example of Application}
The information $G p_e \leq g$ could be exchanged between the TSOs on a regular basis, e.g. every hour after the market clearing, or event-based, e.g. when the tie-line flows change substantially. The exchange could happen bilaterally between the TSOs or over a centralized data exchange. It should be noted, that the TSO of every area controls the amount of his (manual) reserves and redispatch capabilities he shares with the neighboring TSOs by including only the offered reserves in the calculation of the flexibility set.
We briefly sketch below an example how the information could be used.\\

We consider a TSO B that needs to perform a redispatch, e.g. due to inaccurate windfarm forecasts. As TSO B has the information about the possible tie-line flows, he can directly incorporate this information in his redispatch optimization, i.e. the inequality $G p_e \leq g$. In the case of the passive approach, TSO B can guarantee that TSO A will not face congestions as long as TSO B conforms to the deviation limits of the tie-line flows. In the case of the active approach, TSO B can also partially shift the balancing task to TSO A, which will redispatch part of its generation resources. TSO B would have to notify TSO A about the expected tie line flow changes and TSO A would need to be financially compensated for this service. The incurred financial compensation is outside the scope of this paper and is the subject of future work.

\newpage
\section{Case Studies}

\begin{figure}[ht!]
\centering
\includegraphics[width=0.5\linewidth]{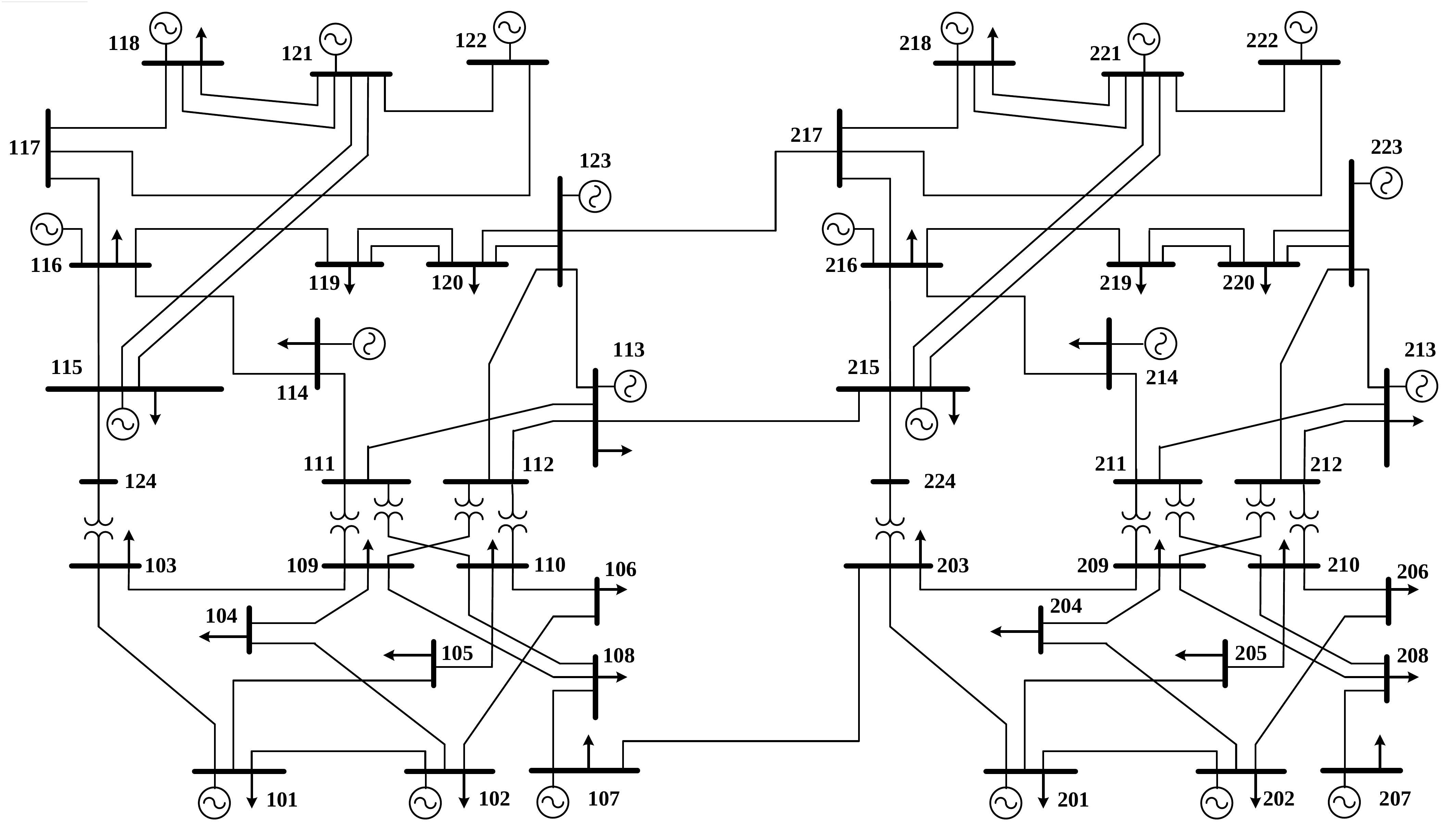}
\caption{IEEE RTS96 - 2 area system. Area A: Buses 1xx. Area B: Buses 2xx}
\label{fig:grid}
\end{figure}

For the case study we use the IEEE RTS96 2 area system as depicted in Fig. \ref{fig:grid}. Area A consists of buses 101 to 123 and area B of buses 201 to 223. The areas are interconnected by three tie-lines and every area is controlled by an individual TSO. For the case study the flexibility set is computed for area A. The loading of the total system is 5700 MW during peak hours and is assumed to drop to approximately 70\% ($=4000MW$) during off-peak hours. The loads are scaled uniformly. For simplicity we assume, that the TSOs are allowed to redispatch every generation unit in the system. All results are in per unit normalized to 100MW.

\subsection{Comparison of Methods}

\begin{figure}[h!]
\centering
\begin{subfigure}{.4\textwidth}
  \centering
\includegraphics[width=.85\linewidth]{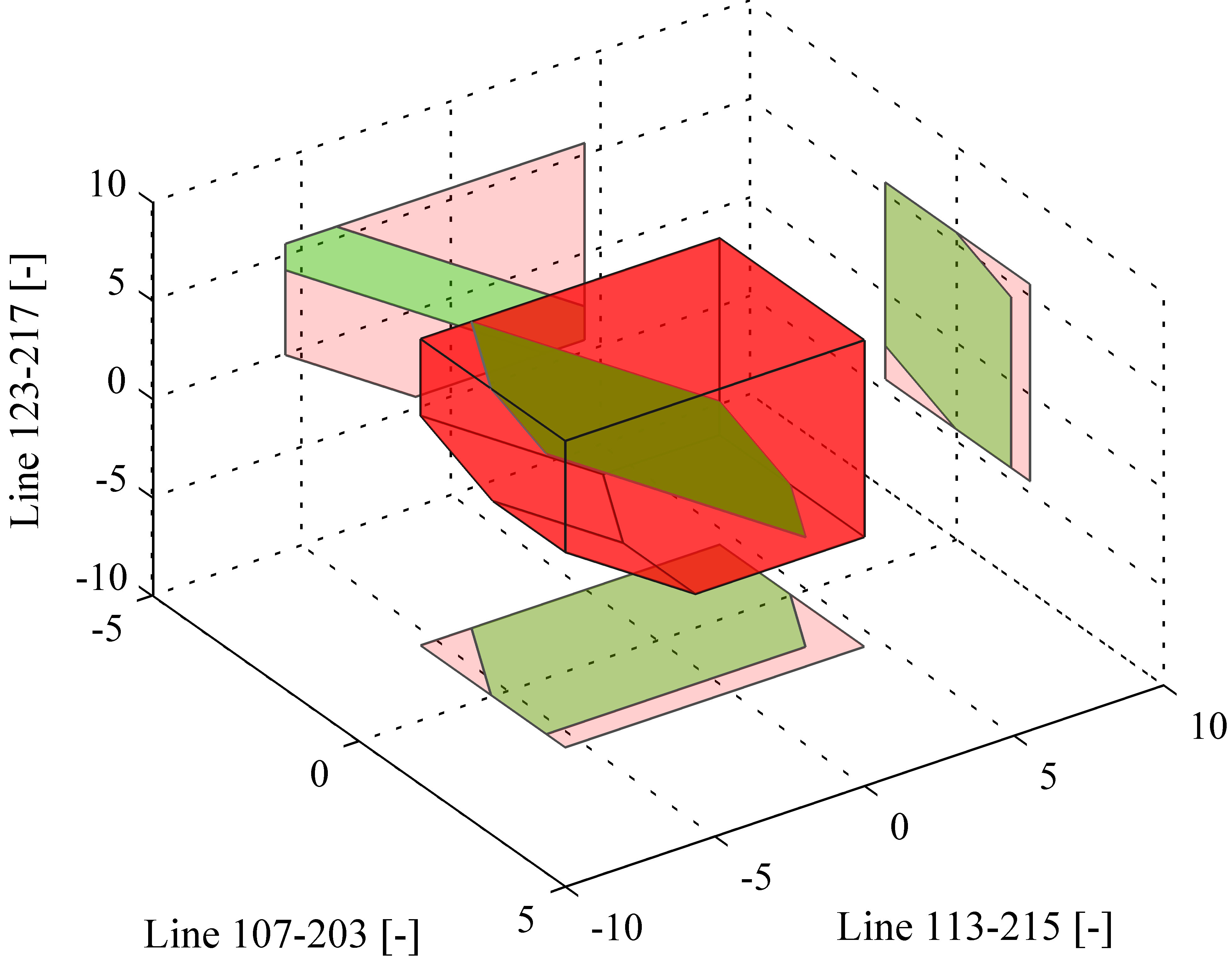}
\caption{Result from the Passive (green) and Active Approach (red) and corresponding projections on planes orthogonal to the axes. The N-1 criterion is not considered.}
\label{fig:N}
\end{subfigure}
\hspace{0.5cm}
\begin{subfigure}{.4\textwidth}
  \centering
\includegraphics[width=.85\linewidth]{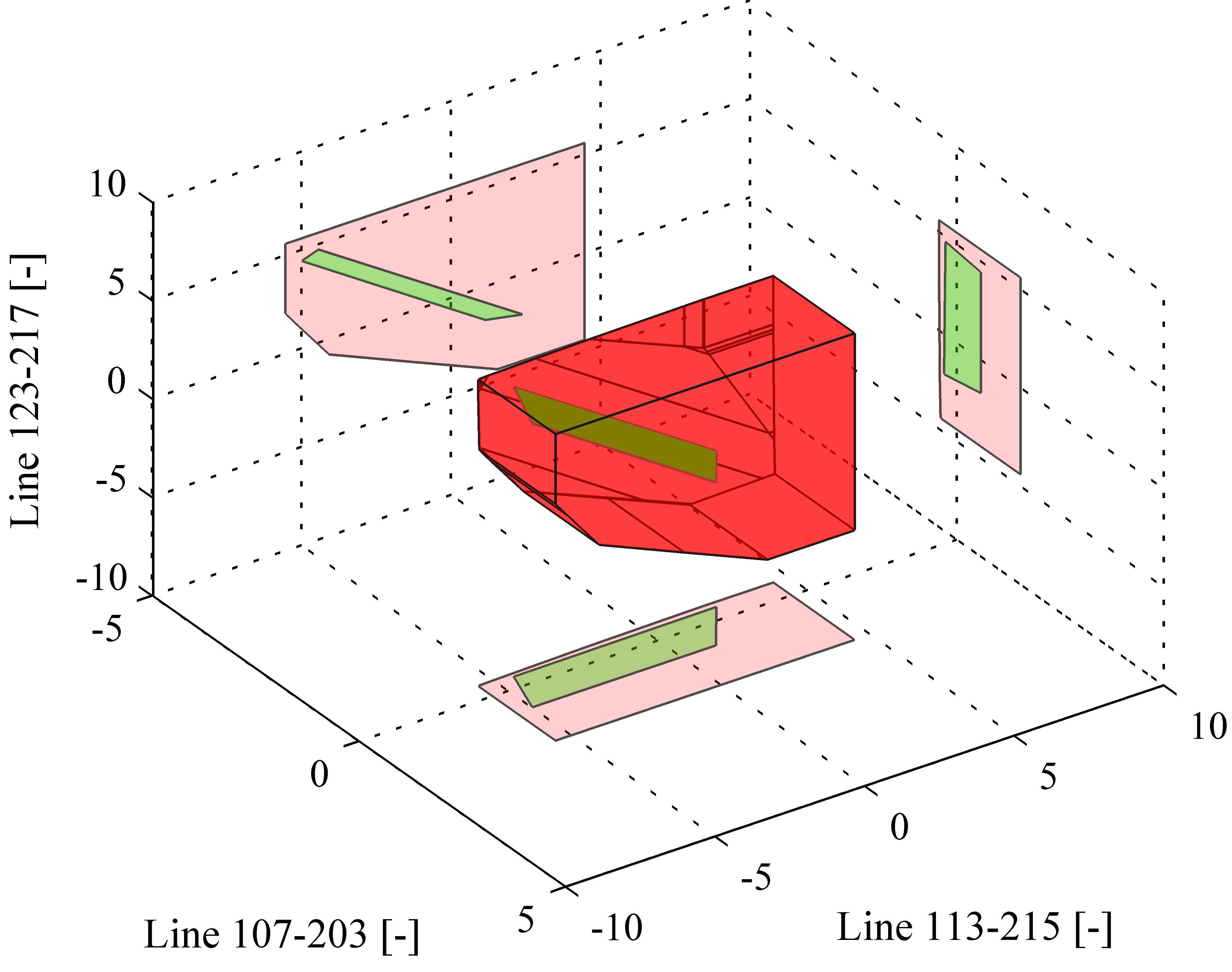}
\caption{Result from the Passive (green) and Active Approach (red) and corresponding projections on planes orthogonal to the axes. The N-1 criterion is considered.}
\label{fig:N1}
\end{subfigure}
\end{figure}

Figs. \ref{fig:N} and \ref{fig:N1} display the feasible sets of the tie-line deviations (during peak load) given by the resulting polytope $G p_e \leq g$. In each figure, the active (red) and the passive (green) approach are compared. Fig. \ref{fig:N} shows the case when the N-1 criterion is not considered and Fig. \ref{fig:N1} when it is considered. The projections of the feasible sets on the planes spanned by the coordinate axes are shown as well. They represent the feasible combinations of two out of the three tie-lines. In the passive approach, only two of three tie-line flows can be chosen as the third tie-line flow has to be such that the total power exchanged with the neighboring areas remains constant. For the active approach, TSO A can support region B using redispatch measures and net energy can be exchanged between the areas. Thus the feasible deviations define a larger set than for the passive approach.

%

\begin{figure}[h!]
\centering
\includegraphics[width=.4\linewidth]{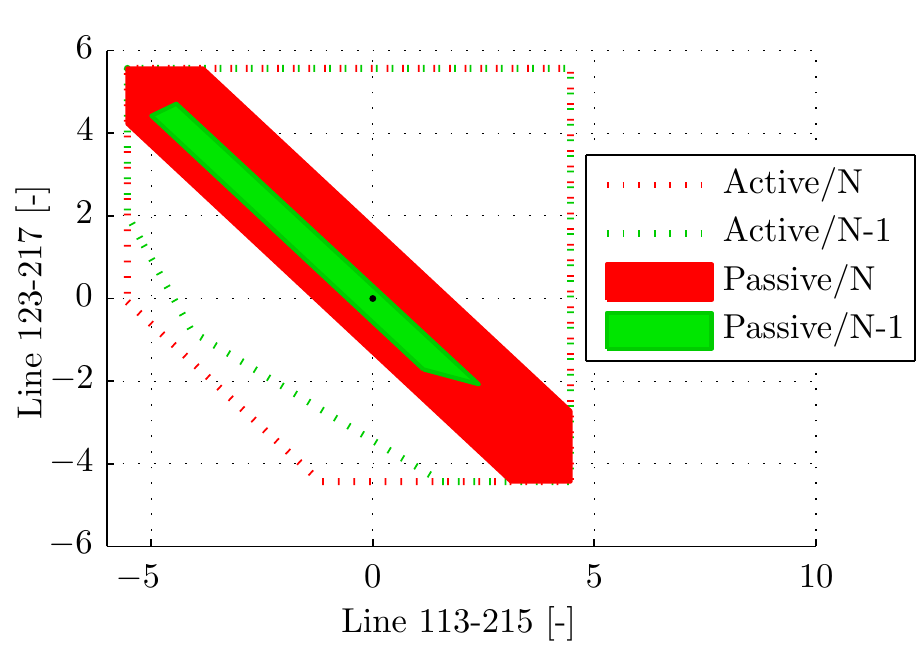}
\caption{Comparison of allowed operational states for two tie-lines.}
\label{fig:projcomp}
\end{figure}

Fig. \ref{fig:projcomp} compares the projections from tie-lines $113-215$ and $123-217$ for both approaches and N-1 and N secure cases. We find the following properties:

\begin{itemize}
\setlength{\itemsep}{1pt}
\setlength{\parskip}{0pt}
\setlength{\parsep}{0pt}
\item $F_{e,N-1} \subseteq F_{e,N}$, using either the passive or active approach: In order to guarantee the N-1 security criterion, a certain flexibility is reserved.
\item $F_{e,passive} \subseteq F_{e,active}$, considering either the N or N-1 secure case: The active approach always adds at least as much flexibility as the passive as generation units can be redispatched.
\item $F_{e,N-1,passive} \subseteq F_{e,N,active}$ holds as logical consequence of the above. But in general it does not hold: $F_{e,N,passive} \subseteq F_{e,N-1,active}$
\item The origin is always contained in all sets, as the current operating state has to be a feasible solution.
\end{itemize}


The illustration of up to three tie-lines is convenient, as they can be represented in 3D graphs. However, even for a larger number of tie-lines, the visualization is feasible. One approach is to project the flexibility sets on relevant combinations of two tie-lines. The visualizations could provide valuable information for example in a dispatch control room of a TSO.
\subsection{Exported Flexibility}
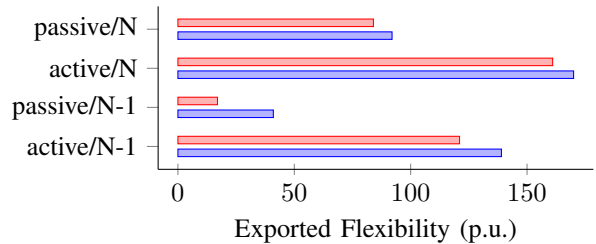
\begin{figure}[h!]
\begin{center}
\begin{tikzpicture}
  \begin{axis}[axis lines=left, axis line style={-},
    xbar, xmin=0,enlarge y limits=0.195, enlarge x limits=0.05,height=3.75cm,
    bar width=0.1cm,width=0.4\textwidth,xlabel={Exported Flexibility (p.u.)},
    symbolic y coords={active/N-1,passive/N-1,active/N,passive/N},]
    \addplot coordinates {(139,active/N-1) (41,passive/N-1) (170,active/N) (92,passive/N)};
    \addplot coordinates {(121,active/N-1) (17,passive/N-1) (161,active/N) (84,passive/N)};
  \end{axis}
\end{tikzpicture}
\caption{Exported Flexibility for peak load (100\%, red bars) and off-peak (70\%, blue bars) load situations.}
\label{fig:exportflex}
\end{center}
\end{figure}

In the introduction, we defined the term {\emph Exported Flexibility}. In order to quantify and compare the amount of flexibility a TSO can provide to its neighbors, the exported flexibility is calculated as the sum of the areas of all projections on all combinations of spanned planes by two tie-lines. For example, considering Fig. \ref{fig:N1} the exported flexibility would be the sum of the areas of the three projections on the $xy$-,$xz$- and $yz$-planes.
Fig. \ref{fig:exportflex} compares the exported flexibility in $p.u.$ for four cases, i.e. passive vs. active approach and N vs. N-1 secure cases.  In general, the available flexibility changes with every change in the load and dispatch. For brevity, we compare it only for two system loadings: 70\%, e.g. during off-peak, and 100\%, e.g. during peak times. We observe that the active approach enables the export of substantially more flexibility than the passive approach, especially in the N-1 case, the increase from passive to active is substantial. In this case the exported flexibility decreases with higher system loadings. Future research could investigate how exported flexibility could be traded as a market product.

\subsection{Tie-Line Utilization}
The remaining transmission capacity (ATC) indicates the limits for transfers between two areas \cite{NERCATC}. The ATC is selected conservatively, considering numerous possible transactions between two areas as well as a number of relevant congestions \cite{Wollenberg2000,Ejebe2000}. This is necessary as the locations of buses involved in a transaction are generally not known. For the active approach, however, the point of injection of reserves is known and thus also the changes in power flows can be anticipated. Therefore it is expected, that the active approach enables a better tie-line utilization than the case with ATC.

In this case study we focus on the utilization of the tie-lines and determine how much flexibility can be imported and exported for the active case compared with the ATC. It should be noted, that the ATC itself does not quantify the amount of balancing that the generation portfolio of region A can provide to region B but quantifies only how much can be transferred between the areas from a grid perspective. The metric determined by the active approach incorporates both, transmission limits and balancing limits.

\noindent The possible power flow deviations on the tie lines that do not exceed the ATC values $ATC_{B \rightarrow A},ATC_{A\rightarrow B}$ are given by the polytope defined by the inequalities:
\begin{equation}
\begin{aligned}
-ATC_{B \rightarrow A} &\leq \sum \Delta P_{L,Tie} \leq ATC_{A\rightarrow B} \\
\Delta\underline{P}_{L,Tie} &\leq \Delta P_{L,Tie} \leq \Delta\overline{P}_{L,Tie}\\
\end{aligned}
\label{eq:ATCPoly}
\end{equation}

$\Delta P_{L,Tie}$ refers to the flow changes on the tie-lines connecting areas A and B. The tie line flow deviation per line cannot exceed the remaining transmission capacity. The ATC from region A to B as well as from B to A is 120MW using the calculation method in \cite{Ejebe2000}.

\begin{figure}[h!]
\centering
\begin{subfigure}{.4\textwidth}
  \centering
  \includegraphics[width=.85\linewidth]{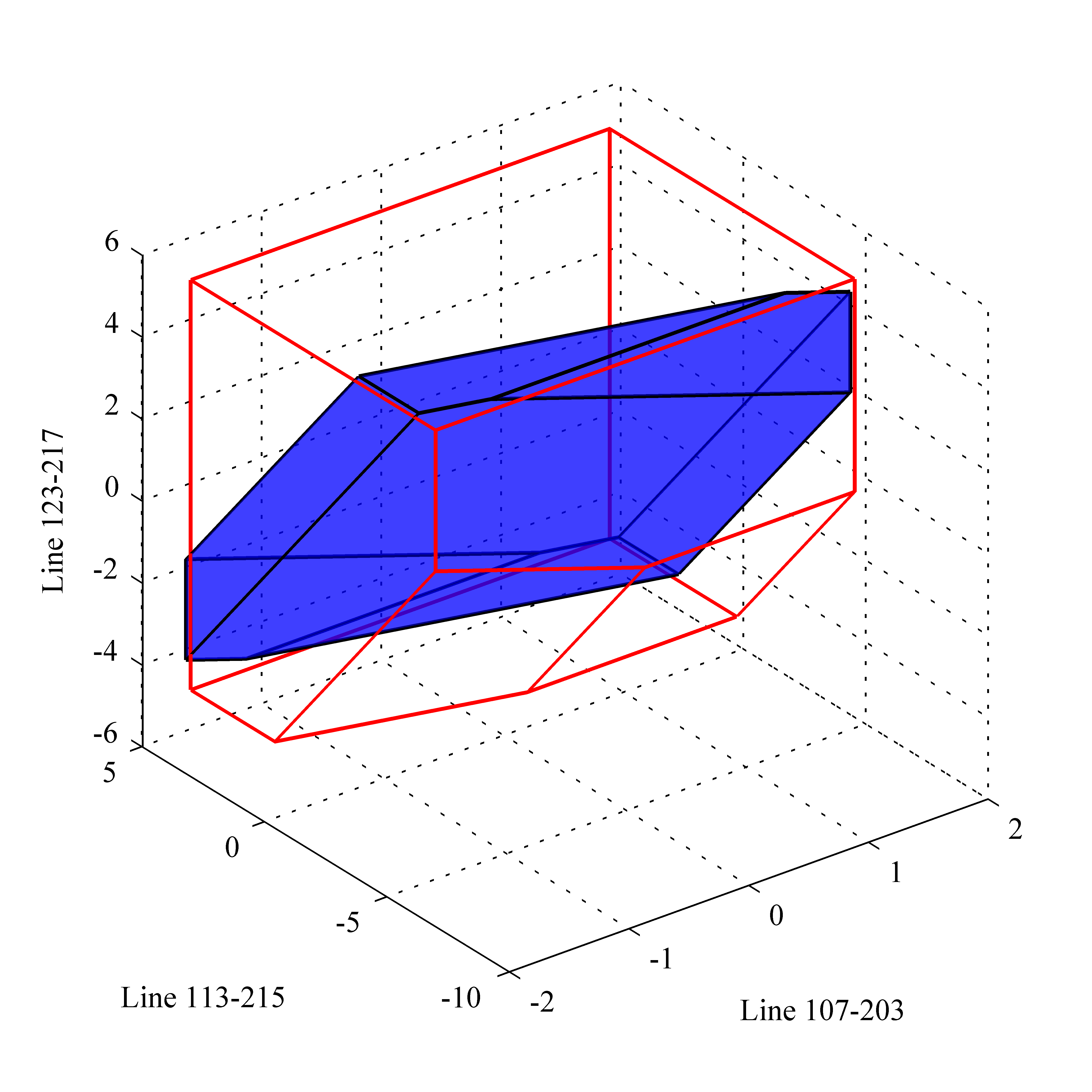}
  \caption{Comparison of ATC (blue) and resulting Polytope of the active approach (red) for full redispatching capabilites in region A.}
  \label{fig:ATCACTIVEALLGEN}
  \end{subfigure}
  \hspace{0.5cm}
  \begin{subfigure}{.4\textwidth}
  \centering
  \includegraphics[width=.85\linewidth]{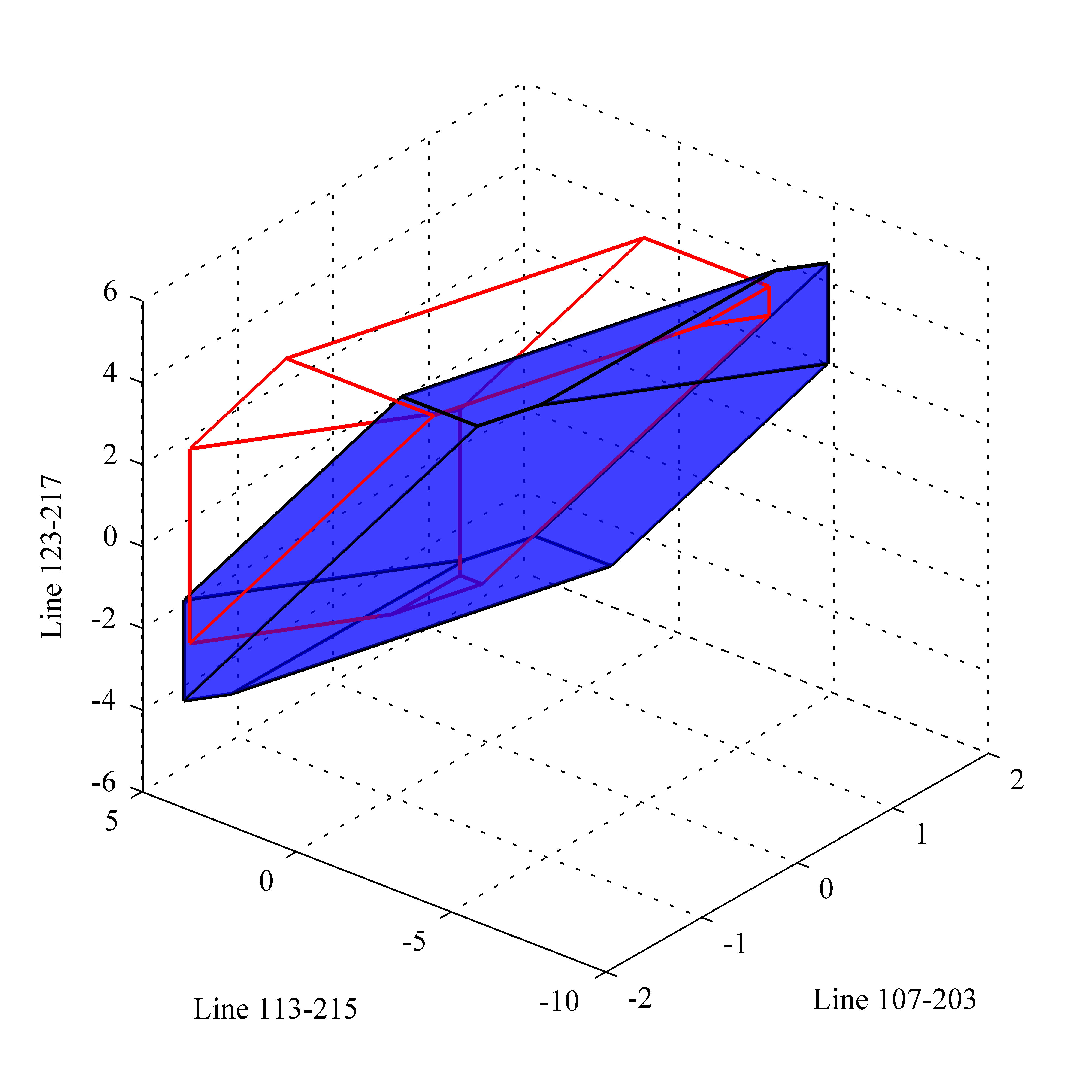}
  \caption{Comparison of ATC and resulting Polytope of the active approach for limited flexibility available in region A. Parts, where ATC polytope is exceeding the area of the active approach corresponds to combinations, where sufficient transfer capacity but insufficient operational flexibility is available in region A.}
  \label{fig:ATCACTIVELASTGEN50}
  \end{subfigure}
\end{figure}

We consider two cases for the availability of flexibility in control area A:
\begin{itemize}
\item TSO A has full access to all generators in area A and can redispatch them. In other words, after the market clearing the remaining up and down capacities of the generation units serve as reserves to TSO A (Fig. \ref{fig:ATCACTIVEALLGEN}).
\item TSO A uses only limited amount of the generating capacity in area A. We assume that TSO A is willing to use the manual reserves of only a single generator in area A to support control area B (Fig. \ref{fig:ATCACTIVELASTGEN50}).
\end{itemize}

Comparing the polytopes resulting from the active approach for the two cases with the polytope for the ATC as shown in Figs. \ref{fig:ATCACTIVEALLGEN} and \ref{fig:ATCACTIVELASTGEN50}, we observe that in the first case, substantially larger deviations are possible when the active approach is used. But also in the case of limited flexibility in area A, some tie-line flow deviations can be alleviated, that would exceed the calculated ATC. The reason for the improved utilization is the knowledge of the amount and location of available flexibility and the corresponding influence on the power flows a TSO has for his control area.
We can conclude, that the active approach enables a less conservative operation of the tie-lines without reducing the security. But the results of the active approach also depend on the availability and location of flexibility in region A.

It should be noted, that in the parts where the ATC polytope is exceeding the area of the active approach, sufficient ATC would be available for transfers between A and B, but the needed redispatch capabilities in control area A would not be available.

\subsection{Maximum Nodal Variations}

In this section, we investigate how the proposed method can mitigate the local deviations arising from for example forecast errors of fluctuating renewable energy sources such as wind or PV. We therefore compare the maximum allowed disturbance of every bus in the control area B for the following three cases:

\begin{itemize}
\item Passive approach: How large can the disturbance in area B be, without causing a congestion in the neighboring region A?
\item Active approach: How large can the disturbance be, when the TSO A helps to compensate the deviation using his procured reserves in area A?
\item ATC: For comparison we also consider the case, where control area B can compensate its deviation not only with its own reserves but also has access to reserves in control area A. For the import/export, the ATC has to be respected.
\end{itemize}

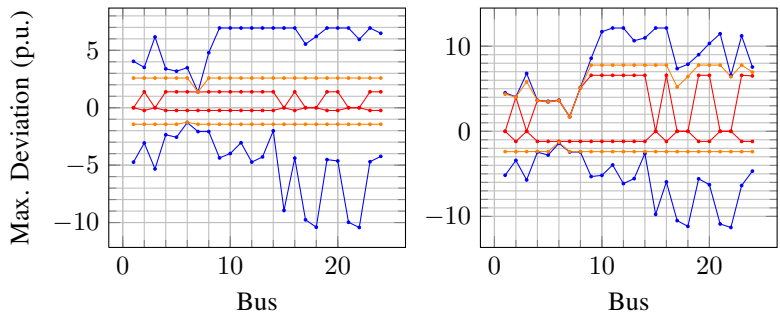
\begin{figure}[ht!]
\begin{center}
\begin{tikzpicture}
	\begin{axis}[width=0.3\textwidth,xlabel=Bus,ylabel=Max. Deviation (p.u.),minor xtick={0,2,...,20},minor ytick = {-10,-9,...,10},grid=both]
	\addplot[color=blue,mark=*, mark size = .5] coordinates {
(1,4.0407)
(2,3.5176)
(3,6.1625)
(4,3.3896)
(5,3.1845)
(6,3.4831)
(7,1.3734)
(8,4.7953)
(9,6.9439)
(10,6.9439)
(11,6.9439)
(12,6.9439)
(13,6.9439)
(14,6.9439)
(15,6.9439)
(16,6.9439)
(17,5.5415)
(18,6.2127)
(19,6.9439)
(20,6.9439)
(21,6.9439)
(22,5.9521)
(23,6.9439)
(24,6.4869)
	};
\addplot[color=red,mark=*, mark size = .5] coordinates {
(1,0)
(2,1.387)
(3,0)
(4,1.387)
(5,1.387)
(6,1.387)
(7,1.3734)
(8,1.387)
(9,1.387)
(10,1.387)
(11,1.387)
(12,1.387)
(13,1.387)
(14,1.387)
(15,0)
(16,1.387)
(17,0)
(18,0)
(19,1.387)
(20,1.387)
(21,0)
(22,0)
(23,1.387)
(24,1.387)
};
\addplot[color=orange,mark=*, mark size = .5] coordinates {
(1,2.587)
(2,2.587)
(3,2.587)
(4,2.587)
(5,2.587)
(6,2.587)
(7,1.3734)
(8,2.587)
(9,2.587)
(10,2.587)
(11,2.587)
(12,2.587)
(13,2.587)
(14,2.587)
(15,2.587)
(16,2.587)
(17,2.587)
(18,2.587)
(19,2.587)
(20,2.587)
(21,2.587)
(22,2.587)
(23,2.587)
(24,2.587)};
\addplot[color=blue,mark=*, mark size = .5] coordinates {
(1,-4.7364)
(2,-3.072)
(3,-5.3271)
(4,-2.3554)
(5,-2.5671)
(6,-1.2903)
(7,-2.0753)
(8,-2.0753)
(9,-4.3697)
(10,-3.99)
(11,-3.0478)
(12,-4.7347)
(13,-4.2837)
(14,-2.0117)
(15,-8.9382)
(16,-4.389)
(17,-9.7604)
(18,-10.4056)
(19,-4.5179)
(20,-4.6345)
(21,-9.9733)
(22,-10.4294)
(23,-4.7007)
(24,-4.2352)
	};
\addplot[color=red,mark=*, mark size = .5] coordinates {
(1,0)
(2,-0.23801)
(3,0)
(4,-0.23801)
(5,-0.23801)
(6,-0.23801)
(7,-0.23801)
(8,-0.23801)
(9,-0.23801)
(10,-0.23801)
(11,-0.23801)
(12,-0.23801)
(13,-0.23801)
(14,-0.23801)
(15,0)
(16,-0.23801)
(17,0)
(18,0)
(19,-0.23801)
(20,-0.23801)
(21,0)
(22,0)
(23,-0.23801)
(24,-0.23801)};

\addplot[color=orange,mark=*, mark size = .5] coordinates {
(1,-1.438)
(2,-1.438)
(3,-1.438)
(4,-1.438)
(5,-1.438)
(6,-1.2647)
(7,-1.438)
(8,-1.438)
(9,-1.438)
(10,-1.438)
(11,-1.438)
(12,-1.438)
(13,-1.438)
(14,-1.438)
(15,-1.438)
(16,-1.438)
(17,-1.438)
(18,-1.438)
(19,-1.438)
(20,-1.438)
(21,-1.438)
(22,-1.438)
(23,-1.438)
(24,-1.438)};
	\end{axis}
\end{tikzpicture}
\begin{tikzpicture}
	\begin{axis}[width=0.3\textwidth,xlabel=Bus,minor xtick={0,2,...,20},minor ytick = {-10,-9,...,10},grid=both]
	\addplot[color=blue,mark=*, mark size = .5] coordinates {
(1,4.5232)
(2,4.0441)
(3,6.7984)
(4,3.6464)
(5,3.5297)
(6,3.6644)
(7,1.7158)
(8,5.158)
(9,8.5706)
(10,11.7036)
(11,12.1378)
(12,12.1378)
(13,10.6464)
(14,10.9705)
(15,12.1378)
(16,12.1378)
(17,7.3642)
(18,7.8794)
(19,9.0033)
(20,10.321)
(21,11.4626)
(22,6.5569)
(23,11.2145)
(24,7.5551)
	};
\addplot[color=red,mark=*, mark size = .5] coordinates {
(1,0)
(2,4.0018)
(3,0)
(4,3.62)
(5,3.4404)
(6,3.6162)
(7,1.7158)
(8,5.093)
(9,6.5808)
(10,6.5808)
(11,6.5808)
(12,6.5808)
(13,6.5808)
(14,6.5808)
(15,0)
(16,6.5808)
(17,0)
(18,0)
(19,6.5808)
(20,6.5808)
(21,0)
(22,0)
(23,6.5808)
(24,6.5075)};

\addplot[color=orange,mark=*, mark size = .5] coordinates {
(1,4.3661)
(2,4.0273)
(3,5.812)
(4,3.6312)
(5,3.4758)
(6,3.6366)
(7,1.7158)
(8,5.1174)
(9,7.7808)
(10,7.7808)
(11,7.7808)
(12,7.7808)
(13,7.7808)
(14,7.7808)
(15,7.7808)
(16,7.7808)
(17,5.2087)
(18,6.4341)
(19,7.7808)
(20,7.7808)
(21,7.7808)
(22,6.4064)
(23,7.7808)
(24,6.9553)};
\addplot[color=blue,mark=*, mark size = 0.5] coordinates {
(1,-5.164)
(2,-3.422)
(3,-5.7126)
(4,-2.4545)
(5,-2.79)
(6,-1.4108)
(7,-2.4367)
(8,-2.4367)
(9,-5.3185)
(10,-5.1779)
(11,-3.9552)
(12,-6.1445)
(13,-5.5591)
(14,-2.6107)
(15,-9.7583)
(16,-5.9476)
(17,-10.4938)
(18,-11.173)
(19,-5.5809)
(20,-6.2802)
(21,-10.8884)
(22,-11.3036)
(23,-6.3698)
(24,-4.6811)
	};
\addplot[color=red,mark=*, mark size = .5] coordinates {
(1,0)
(2,-1.19)
(3,0)
(4,-1.19)
(5,-1.19)
(6,-1.19)
(7,-1.19)
(8,-1.19)
(9,-1.19)
(10,-1.19)
(11,-1.19)
(12,-1.19)
(13,-1.19)
(14,-1.19)
(15,0)
(16,-1.19)
(17,0)
(18,0)
(19,-1.19)
(20,-1.19)
(21,0)
(22,0)
(23,-1.19)
(24,-1.19)};

\addplot[color=orange,mark=*, mark size = .5] coordinates {
(1,-2.39)
(2,-2.39)
(3,-2.39)
(4,-2.368)
(5,-2.39)
(6,-1.3042)
(7,-2.39)
(8,-2.39)
(9,-2.39)
(10,-2.39)
(11,-2.39)
(12,-2.39)
(13,-2.39)
(14,-2.39)
(15,-2.39)
(16,-2.39)
(17,-2.39)
(18,-2.39)
(19,-2.39)
(20,-2.39)
(21,-2.39)
(22,-2.39)
(23,-2.39)
(24,-2.39)};
\end{axis}
\end{tikzpicture}
\caption{Maximal possible deviations at buses in region B for active approach (blue) and passive approach (red) and for comparison the ATC case (orange). The figures on the left are for a Region B with low flexibility (5\%), the figure on the right for a region B with high flexibility (25\%). Positive values correspond to positive deviations, i.e. additional injections, negative values to negative deviations, i.e. additional consumption.}
\label{fig:nodalvar}
\end{center}
\end{figure}

\noindent For simplicity, we assume that the reserves in control area B are given as a percentage of the dispatched units in B, e.g. if the reserves are 5\%, a generator dispatched with 100MW provides $\pm$5MW of reserves. In region A, the generation units can be fully redispatched.
We consider two cases: in the first case, the reserves are 5\%, which corresponds to  $\pm$142.5MW, and in the second case assume a highly flexible system with 25\%, which corresponds to $\pm$712.5MW of reserves in total. The ATC from region A to B as well as from B to A is 120MW using the calculation method in \cite{Ejebe2000}. The goal is not to provide absolute values, but rather compare the relative changes for control areas with high and with low inherent flexibility. For simplicity, the N-1 criterion is not considered for the methods and the ATC calculation.
Fig. \ref{fig:nodalvar} shows the resulting maximal deviations per bus in area B. The possible deviations are different depending on the bus, but the deviations can become the largest in the case of the active approach. It is obvious, that the deviations can be larger when two areas share their reserves compared to the passive approach. If area B is more flexible, larger deviations can be balanced and the dependence on TSO A is reduced. 

\section{Conclusions and Outlook}
\noindent This paper presented a framework for efficiently characterizing and coordinating available operational flexibility between TSOs. %
We therefore introduce the term ``exported flexibility'', which measures the flexibility that one TSO can offer to its neighbors. The information about this flexibility is based on computational geometry and results in a linear matrix inequality that bounds all the feasible tie-line flow deviations.
We distinguish between a passive approach, where TSOs are not expected to deviate from their generation schedule to relieve congestions in neighboring areas, and an active approach, where corrective measures from neighboring TSOs are considered. The N-1 security criterion can also be included.

Case studies compare the proposed approaches for different system loadings. We show that the active approach enables the export of substantially more flexibility. Especially in the case where the N-1 criterion is considered, the increase in the ``exported flexibility'' is between 240\% - 600\% depending on the system loading. The deviations the system can cope with is substantially larger if flexibility is shared between two areas and therefore allows the incorporation of more intermittent energy sources. A comparison with the ATC also shows, that the active approach allows to improve the tie-line utilization.

Future work will focus on the incorporation of ramping constraints imposed by the generation units as well as a pricing scheme for redispatch measures. Further, the generalization to more than two areas will be developed.

\footnotesize
\section*{Acknowledgement}
This research was carried out within the project Balancing Power in the European System (BPES). Financial support by the Swiss Federal Office of Energy (SFOE) and Swissgrid is gratefully acknowledged.

\newpage
\bibliographystyle{IEEEtran}
\bibliography{Bibliography}

\end{document}